\documentclass[reprint,amsmath,amssymb,aps]{revtex4-2}
\usepackage{amsmath,amsfonts,upgreek,amsbsy}
\usepackage{graphicx,xcolor}

\begin{document}

% Title
\title{Pleiotropy enables specific and accurate signaling in the presence of ligand cross talk}

% Use letters for affiliations, numbers to show equal authorship (if applicable) and to indicate the corresponding author

\iffalse
\author[1\authfn{1}]{Duncan Kirby}
\author[1\authfn{1}]{Jeremy Rothschild}
\author[1\authfn{1}]{Matthew Smart}
\author[1,2]{Anton Zilman}

\affil[1]{Physics Department, University of Toronto, 60 Saint George st, Toronto, ON M5S1A7, Canada}
\affil[2]{Institute for Biomaterials and Bioengineering, University of Toronto, 164 College St, Toronto, ON M5S 3G9, Canada}

\corr{zilmana@physics.utoronto.ca}{A.Z.}

\contrib[\authfn{1}]{Alphabetical order. D.K, J.R, and M.S contributed equally to this work.}

\fi

\author{Duncan Kirby$^a$}
\author{Jeremy Rothschild$^a$}%
\author{Matthew Smart$^a$}%
\author{Anton Zilman$^{a,b}$}%
\email{zilmana@physics.utoronto.ca}
\affiliation{$^a$University of Toronto, 60 Saint George st, Toronto, ON M5S1A7, Canada}
\affiliation{$^b$Institute for Biomaterials and Bioengineering, University of Toronto, 164 College St , Toronto, ON M5S 3G9, Canada}

\begin{abstract}
Living cells sense their environment through the binding of extra-cellular molecular ligands to cell surface receptors. Puzzlingly, vast numbers of signaling pathways exhibit a high degree of cross talk between different signals whereby different ligands act through the same receptor or shared components downstream. It remains unclear how a cell can accurately process information from the environment in such cross-wired pathways. We show that a feature which commonly accompanies cross talk - signaling pleiotropy (the ability of a receptor to produce multiple outputs) - offers a solution to the cross talk problem. In a minimal model we show that a single pleiotropic receptor can simultaneously identify and accurately sense the concentrations of arbitrary unknown ligands present individually or in a mixture. We calculate the fundamental limits of the signaling specificity and accuracy of such signaling schemes. The model serves as an elementary ``building block'' towards understanding more complex cross-wired receptor-ligand signaling networks.
\end{abstract}

\maketitle

\section{Introduction}
\label{sec:intro}
Receptor signaling via soluble ligand molecules enables living cells
to communicate with each other and with their environment, and is
the main mode of multicellular coordination in the immune, nervous,
endocrine and other systems, as well as in complex populations of
micro-organisms. In a typical signaling pathway, binding of a ligand
to a cell surface receptor activates a cascade of intracellular events
that eventually lead to responses such as cellular differentiation
\citep{Laurenti2018,Crossman2018,geissmann2010development,Lodish2000},
phenotypic change \citep{Altan-Bonnet2005,Bagnall2018,Touzot2014,Lodish2000},
or a change in cellular motility \citep{Rappel2008,Berg1972,wang2011signaling,wong2018exploring,micali2016bacterial}.
For reliable and precise communication, receptor signaling often needs
to be specific, accurate, rapid and robust to molecular noise and
cellular heterogeneity \citep{Berg1977,Bialek2005,TenWolde2016,Cui2018,Siggia2013,Francois2013,bialek2012biophysics,ladbury2012noise}.
However, fundamental physical constraints often place these different
signaling goals at odds with each other \citep{Suderman2017,Fathi2016,Francois2013,Francois2016,Siggia2013,MehtaMora2014},
and different signaling pathways have evolved to optimize different
aspects of information transmission such as molecular specificity \citep{McKeithan1995,Francois2013,Fathi2016},
sensitivity \citep{Song2006,Altan-Bonnet2005,Francois2013}, accuracy of concentration sensing
\citep{Berg1972,gradientsensingEndres2009,Ellison2016,Mora2015},
and speed \citep{MehtaMora2014,Murugan2012,Banerjee2017,Tran2018}.

Puzzlingly, receptor signaling pathways frequently exhibit a high
degree of cross talk whereby multiple ligands act through shared cell
surface receptors or downstream signaling components \citep{Garbers2012,Basak2008,Rawlings2004,Thomas2011,Antebi2017,Klinke2012,McClean2007,Moraga2014,Thomas2011,Jetka2018,Kholodenko2010,chatterjee2010pairwise,Schreiber2017}.
Another puzzling feature that commonly accompanies cross talk is receptor pleiotropy - the ability of a receptor to produce more than one type of output. This combination of features commonly results in ``hourglass''
shaped input-output networks \citep{Jetka2018,Moraga2014,Ozaki2002,Lin1995,moggs2001estrogen,tagaya199615}.
This challenges the classical ``one ligand/one signal'' paradigm \citep{Lodish2000}, while raising the question of how signaling pathways
are able to effectively transmit information under such conditions \citep{Vert2011, Carballo-Pacheco2019,Mora2015,Singh2017,Singh2020,Fathi2016,Jetka2018,komorowski2019limited,zinkle2018threshold}.

Ligand-receptor cross talk poses a fundamental problem for effective signal transmission which  can be illustrated through an example of the conflict between two goals: 1) identification of the ligand out of many others that bind the same receptors in order to produce a specific response (which we denote as ``specificity") and 2) quantification of the amount of the ligand in order to accurately respond to different concentrations (which we denote as ``accuracy"). In a classical view of ligand-receptor binding, the signaling response is dictated by
the average receptor occupancy $P=(c/K_{d})/(1+c/K_{d})$, where $c$ is the ligand concentration and $K_{d}$ is the equilibrium dissociation constant of the ligand-receptor binding \citep{phillips2012physical}.
The ligand concentration and the dissociation constant enter into this expression only through their ratio, $c/K_{d}$, and identical receptor occupancies can be realized by a weakly binding ligand present at a high concentration, or a strongly binding one at a low concentration.
Hence, based on the receptor occupancy alone, it is generally impossible
to unambiguously determine which ligand is bound to the receptor while also accurately quantifying its concentration (see \ref{fig:mode_1_fig}.
This example is a manifestation of a more general inference problem that arises in the presence of cross talk: the difficulty of unambiguously inferring multiple input variables - in this  case the ligand concentration (``quantity'') and its identity measured as receptor binding affinity (``quality'') - from one output variable (in this case receptor occupancy).
This problem is exacerbated when multiple ligands that bind the same receptor are present simultaneously in a mixture, as the number of unknown ligand identities and concentrations increases.
It remains unclear how the cell can unambiguously determine the composition and the concentrations of the ligands in the mixture.

Several recent works addressed some aspects of this problem. In particular, \citep{Mora2015,Carballo-Pacheco2019,Singh2017,Singh2020,Jetka2018}
focused on the effects of cross talk on the accuracy of sensing the concentration of a cognate ligand in the presence of non-specific
ligands. Such problems commonly arise in the context of cellular chemotaxis
driven by concentration gradients of food or chemoattractants. Using an extension of the classical Berg-Purcell framework
\citep{Berg1977,TenWolde2016,Bialek2005,Endres2009,Skoge2011},
it was shown that, given sufficient separation between the affinities
of the specific and non-specific ligands, detection and accurate sensing
of the concentration of the high-affinity ligand is possible even
if it is outnumbered by the low-affinity ligand \citep{Singh2017,Mora2015}. Furthermore, under certain
conditions the cell is able to determine the concentration
of the non-specific ligands as well \citep{Singh2017}. In further
work, it was also shown that the accuracy can in some cases be increased by a more complicated ligand-receptor
network that includes two cross-wired receptors \citep{Carballo-Pacheco2019}.
The results of these works largely rely on several  important assumptions: 1) only two ligand types (specific and non-specific) can bind the same receptor, 2) identity of the cognate ligand, as expressed via its binding affinity
to the receptor, is known and 3) the inference of the ligand concentration is based on the whole
sequence of ligand-receptor binding and unbinding times.

In a related set of problems, inspired by T cell receptor
(TCR) signaling, the cell needs to efficiently filter out the weak affinity ``self'' (non-specific) ligands but to sensitively respond to even very low concentrations of the strong affinity ``non-self'' (specific) ligand.
One proposed solution, relying on the ``adaptive sorting'' modification of the classical kinetic proofreading (KPR) scheme, enables ``absolute discrimination'' between different ligands based on their affinities, unconfounded by their concentrations \citep{Francois2013,Francois2016}.
Kinetically different, but conceptually similar mechanisms are involved in the absolute ligand discrimination by dimeric receptors \citep{Fathi2016}.
In a different approach to the same problem, \cite{Siggia2013,Mora2015} studied the probability of detection of a cognate ligand present at a low concentration over a background of ``wrong'' ligands.
However, these works focused on the sensitive detection of the ligand presence rather than on the identification of different ligands.

In this paper we consider a general problem of specific (ligand identification) and accurate (concentration measurement) sensing  in the presence of cross talk in signaling pathways with multiple ligands acting through a single shared receptor.
This problem is motivated by the observation that in many signaling systems cells are capable of  providing substantially different responses to multiple different ligands acting through the same pathway (specificity/identification) while maintaining dose response sensitivity for each of them (accuracy/quantification), even if the ligands are present in complex mixtures. This scenario appears in a number of signaling pathways such as cytokine and chemokine signaling, some aspects of T cell response, G-protein coupled receptor (GPCR) signaling and  others\citep{Wootten2018,Moraga2014,Sokol2015,zinkle2018threshold}.
The ability to specifically and accurately respond to different cytokine combinations is necessary for cells to respond to different physiological situations which are encoded by different combinations of signaling molecules \cite{Antebi2017,tkach2014cracking,li2019communication}.
This raises the question of how cells are able to sense mixed signals both specifically and accurately in the presence of cross talk \citep{Moraga2014,Jetka2018}.

We show that signaling pleiotropy - the ability of the receptor to produce several different signals in response to ligand binding - provides a solution to the specificity-accuracy problem both when the ligands are present alone or in mixtures.
We introduce a minimal, biologically motivated model of a pleiotropic receptor capable of binding a large number of ligands, and show that it can unambiguously determine both the receptor binding affinities and the concentrations
of an arbitrary number of ligands. Furthermore, we calculate the fundamental limits
on the specificity and accuracy of such sensing in the presence of noise both at the receptor level and downstream.

We further show that signaling pleiotropy enables specific and accurate sensing of the affinities and concentrations when two ligands are present in a mixture, demonstrating that the receptor signaling scheme discussed here can serve as a ``building block'' for the general problem of signaling cross talk in mixtures of multiple ligands.

The paper is structured as follows. In section \ref{subsec:non-pleio} we define the mathematical framework, formulate the problem of the specificity-accuracy trade-off for a non-pleiotropic receptor and introduce the mathematical definitions of specificity and accuracy.
In section \ref{subsec:pleiotropic-receptor} we introduce the
pleiotropic receptor, show that it resolves the specificity-accuracy
problem, and calculate the fundamental limits on the sensing specificity and accuracy. In section \ref{subsec:ligands-mixture} we show how our model resolves the cross talk problem in more complex cases where multiple ligands are present simultaneously. We conclude
with a discussion and possible generalizations in section \ref{sec:sum-dis}.
\section{Results}
\label{sec:results}
\subsection{Non-pleiotropic receptor}
\label{subsec:non-pleio}
The accuracy of concentration sensing by a molecularly specific receptor that binds a ligand of one type has been investigated in a large number of works \citep{nelson2014physical,Endres2009,TenWolde2016,Kaizu2014,Singh2017,Mora2015}, starting with the pioneering work of Berg and Purcell \citep{Berg1977}. In this section, we extend the framework developed in these works to address the problem of specificity - the ability of the receptor to distinguish between different molecularly distinct ligands.

\begin{figure*}[!h]
\centering
\includegraphics[width=12.9 cm]{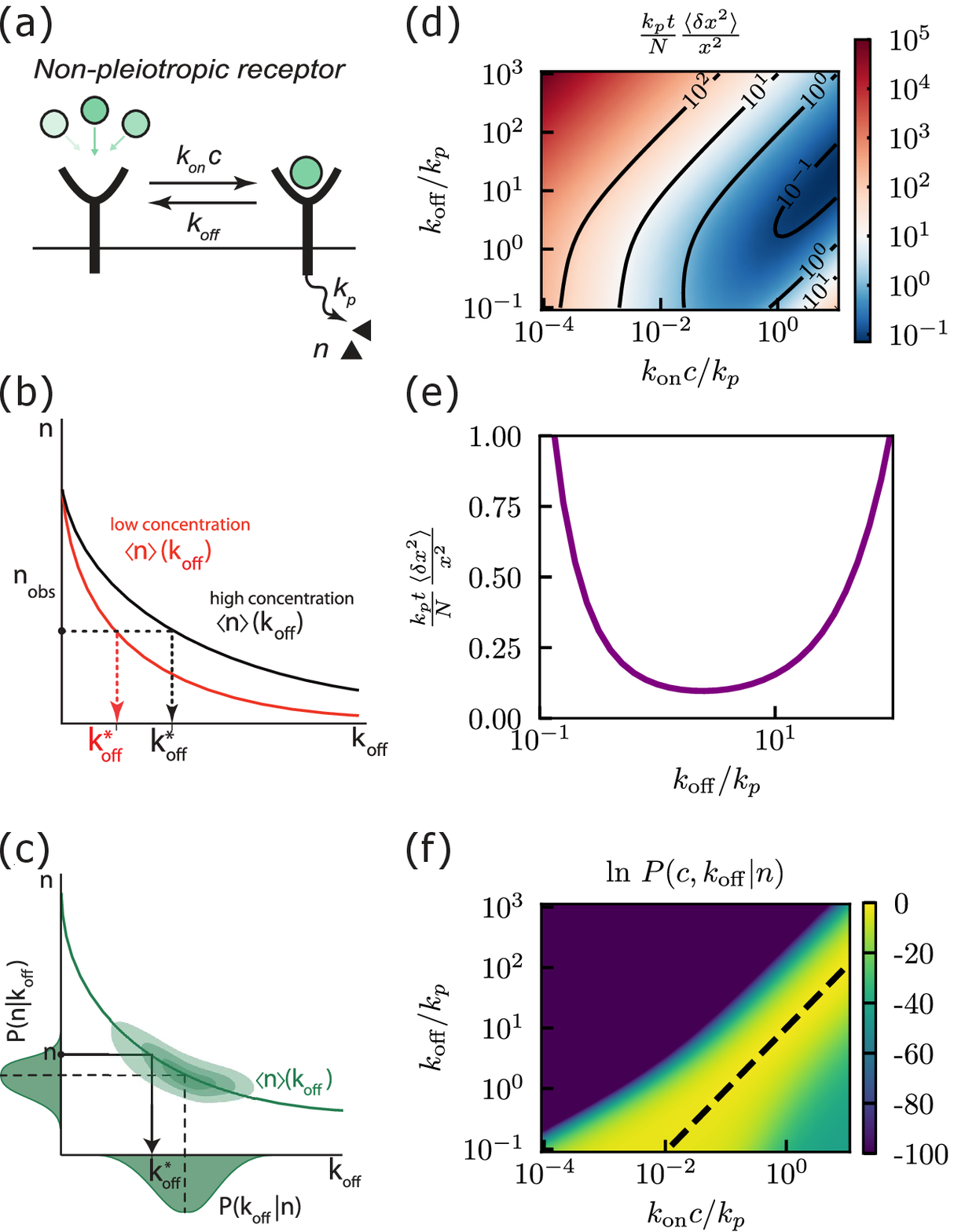}
\caption{\textit{Specificity and accuracy for a non-pleiotropic receptor.}
(a) Different ligands (shown in different colors) with
distinct $k_{\text{off}}$ can bind the receptor. When bound by a
ligand, the receptor produces downstream signaling molecules $n$
at a rate $k_{p}$. (b) An illustration of the fundamental
specificity-accuracy problem. The same number of signaling molecules
$n_{\text{obs}}$ can be produced by a low concentration of a
strongly binding ligand as a high concentration of
a weakly binding ligand. See text. (c) Fluctuations in $n$ resulting
from the stochasticity of the binding-unbinding and production events, described
by the probability distribution, $P(n|k_{\text{off}})$, lead to
an uncertainty in the estimate $k_{\text{off}}^{*}$, encapsulated in the distribution
$P(k_{\text{off}}|n).$ The width of $P(k_{\text{off}}|n)$ is a measure of the uncertainty
in the estimate. (d) Squared relative error of the estimate
of $x$ as a function of the dimensionless ligand concentration and
unbinding rate, scaled by $k_{p}t/N$ with $k_{p}t=10^{2}$ and $N=10^{2}$; see Eq. (\ref{delta-x-from-n}). (e) A
cross-section of the heatmap in panel (d), holding ${k}_{\text{on}}c/k_{p}=1$.
(f) The posterior probability $P(c,k_{\text{off}}|n)$ does
not have a defined maximum in $(c,k_{\text{off}})$ space but instead
has a ridge along the line $ (\frac{k_{\text{on}}c}{k_p}) = (\frac{k_{\text{off}}}{k_p})\frac{n}{k_{p}t}\frac{1}{1-n/k_{p}t}$;
 $n=10^{3}$ and $k_{p}t=10^{2}$, illustrating the impossibility
of simultaneous determination of $c$ and $k_{\text{off}}$ by a non-pleiotropic
receptor.}
\label{fig:mode_1_fig}
\end{figure*}

To investigate the general problem of signaling specificity and accuracy,
in this section we introduce a signaling receptor capable of binding
a large number of ligands, as illustrated in Fig. \ref{fig:mode_1_fig}.
Here we confine ourselves to the situation where the receptor is exposed to a single ligand, out of many possible ones that bind the same receptor due to ligand cross talk, at concentration $c$ (see section \ref{subsec:ligands-mixture} for generalization to ligand mixtures).

The identity of a ligand is defined by its binding and
unbinding rates to the receptor, $k_{\text{on}}$ and $k_{\text{off}}$, respectively. 
In general, these rates depend in a non-trivial fashion
on the molecular details of the receptor-ligand interface \citep{Moraga2014,Thomas2011,Schreiber2017}
and its surroundings \citep{TenWolde2016,Faro2017,bromley2001immunological};
for simple monomolecular binding they combine into the equilibrium
dissociation constant $K_{d}\equiv k_{\text{off}}/k_{\text{on}}\propto e^{-\epsilon}$
where $\epsilon$ is the ligand-receptor binding energy \citep{lauffenburger1996receptors}. For simplicity,
we assume that the binding rate constant $k_{\text{on}}$ is independent of
the ligand identity, which is then fully captured by its unbinding
rate $k_{\text{off}}$ (or, alternatively, the dissociation constant $K_d=k_{\text{off}}/k_{\text{on}}$.

While bound by a ligand, a non-pleiotropic receptor produces a single type of downstream
signaling molecule at a rate $k_{p}$, which serves as the readout of
the ligand presence outside the cell. This signaling mechanism is
common to a large number of pathways, e.g. where the active form of the
output molecule is produced via phosphorylation by a receptor-bound
kinase \citep{Lodish2000,Govern2014}. Functionally, this output variable
essentially measures the bound time of the receptor \citep{TenWolde2016}.
The results which follow are trivially extended to $N$ independent copies of the receptor, as explained below.

The state of the system at time $t$ is described by the probability
$P_{i}^{n}(t)$ to be in the occupancy state $i$ ($i=1$ when the
receptor is occupied by the ligand, $0$ otherwise) and have produced
$n$ output molecules by time $t$. The ensemble dynamics of the system
are described by the master equation for the probability $P_{i}^{n}(t)$
\citep{Nemenman2010,Jafarpour2017, pinsky2010introduction}:
\begin{equation}
\begin{aligned}\frac{d}{dt}P_{0}^{n} & =k_{\text{off}}P_{1}^{n}-k_{\text{on}}cP_{0}^{n}\\
\frac{d}{dt}P_{1}^{n} & =k_{\text{on}}cP_{0}^{n}-k_{\text{off}}P_{1}^{n}+k_{p}P_{1}^{n-1}-k_{p}P_{1}^{n}
\end{aligned}
\label{eq-ME-onerec}
\end{equation}

At receptor occupancy steady state, the probability of the receptor being occupied is
$p=x/(1+x)$ where $x=c/K_{d}$. For simplicity and connection to
previous work \citep{Endres2009,Mora2015}, we currently
neglect degradation of the output molecules.

The master equation {(}\ref{eq-ME-onerec}{)} can be solved using
the generating function $G_{i}(s,t)=\sum_{n}s^{n}P_{i}^{n}(t)$. The
dynamics of the vector $\textbf{G}=\left(G_{0}(s,t),G_{1}(s,t)\right)$
are then described by the equation \citep{Nemenman2010,pinsky2010introduction}:
\begin{equation}
\begin{aligned}\frac{d}{dt}\textbf{G}(s,t) & =\pmb{\hat{\text{M}}}\textbf{G}(s,t)\\
\;\;\text{ with}\;\pmb{\hat{\text{M}}} & =\begin{bmatrix}-k_{\text{on}}c & k_{\text{off}}\\
k_{\text{on}}c & -k_{\text{off}}+k_{p}(s-1)
\end{bmatrix},\label{gen-fun-eq}
\end{aligned}
\end{equation}
%\hat{\textbf{M}}
yielding the general solution at time $t$ as $\textbf{G}(s,t)=e^{\hat{\textbf{M}}t}\textbf{G}(s,0)$.
Assuming that the receptor is at steady state at the beginning of
the measurement (defined as $n=0$), the initial condition is $\textbf{P}^{0}=(1-p,p)$
for $n=0$ and $\textbf{P}^{n}=(0,0)$ otherwise, and therefore $\textbf{G}(1,0)=(1-p,p)$.
Similar results can be obtained if the receptor is initially unoccupied
with $\textbf{G}(1,0)=(1,0)$ producing identical results in the limit
of large numbers of binding events, $k_{\text{off}}t\gg1$ which is
the focus of this paper (see Supplemental Material \citep{SM2020} for details).

The mean and the variance of $n$ are calculated as $\langle n\rangle=\sum_{i}\frac{\partial G_{i}(s,t)}{\partial s}|_{s=1}$
and $\langle{\delta n}^{2}\rangle=\sum_{i}\frac{\partial{G}_{i}(s,t)}{\partial s^{2}}|_{s=1}+\langle n\rangle-\langle n\rangle^{2}$.
For $k_{\text{off}}t\gg1$, these are
\begin{equation}
\begin{aligned}\langle n\rangle & =k_{p}t\frac{x}{1+x}\\
\langle{\delta n}^{2}\rangle & =k_{p}t\frac{x}{1+x}+\frac{2k_{p}^{2}t}{k_{\text{off}}}\frac{x}{(1+x)^{3}}.
\label{nonpleio_meann}
\end{aligned}
\end{equation}

In the long-time limit, defined as $\text{min}(k_{\text{off}},k_{p})\gg1/t$,
the probability distribution of $n$, $P(n|c,k_{\text{off}}),$ tends
to a Normal distribution $\mathcal{N}(\langle n\rangle,\langle{\delta n}^{2}\rangle)$,
$P(n|c,k_{\text{off}})=(2\pi\langle{\delta n}^{2}\rangle)^{-1/2}\exp(-(n-\langle n\rangle)^{2}/(2\langle{\delta n}^{2}\rangle))$
(see Supplemental Material \citep{SM2020}).
Physically, these results reflect the fact that the receptor produces
molecules at rate $k_{p}$ while it is occupied (on average $x/(1+x)$
fraction of the time $t$). For the variance, the first term reflects fluctuations in the readout $n$ for fixed bound time, and the second term reflects fluctuations in the bound time; see also Supplementary Material Section D. Note that the variance of $n$ scales
as $k_{p}t$, in accordance with the Central Limit Theorem. 

\textbf{\emph{Specificity and accuracy of the non-pleiotropic receptor.}} The classical problem of accuracy
can be stated as the estimation of the concentration $c$ from the
number of signaling molecules $n$ when $k_{\text{off}}$ is known
- a situation realized when the receptor is highly molecularly specific
and can bind only one ligand type \citep{TenWolde2016,Berg1977}. We
formulate the problem of specificity in a similar manner - as the
estimation of $k_{\text{off}}$ given $n$. Note that this specificity
definition is different from the common measure of specificity as
the ratio of the mean number of sensing molecules produced by different
ligands. However, it is impossible to unambiguously estimate both $c$
and $k_{\text{off}}$ from a single measurement of the output variable $n$ - illustrating
the fundamental specificity-accuracy trade-off - because the same
number of output molecules can be produced by a weaker ligand at higher
concentration as by a stronger ligand at lower concentration. This
point is illustrated in Fig. \ref{fig:mode_1_fig}(b). Within this
scheme, only the combination $x=k_{\text{on}}c/k_{\text{off}}$ can
be inferred from $n$.

One useful framework to interrogate these ideas is statistical inference theory \citep{KayTextCh3}. Here we provide approximate but intuitive derivations; exact numerical derivations are provided in the Supplemental Material \citep{SM2020}. To help illustrate the specificity-accuracy trade-off, we first focus on the determination of the ligand identity through the estimation of $k_{\text{off}}$. Assuming the concentration $c$
is known, an intuitive estimate of $k_{\text{off}}$ given $n$, $k_{\text{off}}^{*}(n)$, is provided via maximization of the likelihood $P(n|k_{\text{off}})$ (or equivalently, the log-likelihood $L=\ln P$) over $k_{\text{off}}$; this procedure is known as maximum likelihood
estimation (MLE) \citep{nelson2014physical,Endres2009}. This is equivalent
to the maximization of the posterior probability $P(k_{\text{off}}|n)=P(n|k_{\text{off}})P(k_{\text{off}})/\int P(n|k_{\text{off}})P(k_{\text{off}})dk_{\text{off}}$
for a uniform prior $P(k_{\text{off}})$ \citep{nelson2014physical}. In the long time limit, the likelihood $P(n|k_{\text{off}})$ is well-approximated
by a Normal distribution, peaked around $\langle n\rangle$. Neglecting the logarithmic terms in $L(n|k_{\text{off}})$, the MLE $k_{\text{off}}^{*}$
is given by the condition $\langle n\rangle(k_{\text{off}}^{*})=n$ \citep{TenWolde2016,KayTextCh3},
yielding
\begin{equation}
\frac{k_{\text{off}}^{*}}{k_{\text{on}}c}=\frac{k_{p}t}{n}-1\label{eq-c-estimate-simple-receptor}
\end{equation}

For a given $n$, expanding the likelihood about $k_{\text{off}}^{*}(n)$ gives,
to lowest order, a Normal distribution whose variance $\delta k_{\text{off}}^{2}=-\left(\frac{\partial^{2}\ln P(n|k_{\text{off}})}{\partial k_{\text{off}}^{2}}\rvert_{k_{\text{off}}^{*}}\right)^{-1}$
is a measure of the uncertainty in the estimate $k_{\text{off}}^{*}$ \citep{nelson2014physical}. Repeating
this over the distribution $P(n|k_{\text{off}})$ of possible outcomes $n$ gives
the average uncertainty $\langle\delta k_{\text{off}}^{2}\rangle=-\left\langle \frac{\partial^{2}\ln P(n|k_{\text{off}})}{\partial k_{\text{off}}^{2}}\right\rangle ^{-1}$;
the quantity $\left\langle -\frac{\partial^{2}\ln P(n|k_{\text{off}})}{\partial k_{\text{off}}^{2}}\right\rangle $
is known as the Fisher Information \citep{KayTextCh3}. In the saddle
point approximation, accurate for sharply peaked likelihoods, this
simplifies to $\langle\delta k_{\text{off}}^{2}\rangle\simeq-\left(\frac{\partial^{2}\ln P(n|k_{\text{off}})|_{k_{\text{off}}=k_{\text{off}}^{*}}}{\partial k_{\text{off}}^{2}}\right)^{-1}\simeq\langle\delta n^{2}\rangle/(\partial\langle n\rangle/\partial k_{\text{off}})^{2}$
\citep{KayTextCh3,nelson2014physical,TenWolde2016} (see Supplemental Material \citep{SM2020}).
This expression has a simple intuitive meaning: fluctuations in $n$
resulting from the stochasticity of binding-unbinding and production
events lead to uncertainty in the estimate $k_{\text{off}}^{*}$, as illustrated
in Fig. \ref{fig:mode_1_fig}(c). Importantly, although heuristically derived here on the basis of the ML estimate, the Fisher Information formalism applies more generally as a lower bound on the estimate error \citep{KayTextCh3}. 

Using the non-dimensionalized quantity $x=k_{\text{on}}c/k_{\text{off}}$, the relative average error (squared) of the estimate becomes
\begin{equation}
\frac{\langle\delta k_{\text{off}}^{2}\rangle}{k_{\text{off}}^{2}}=\frac{\langle\delta x^{2}\rangle}{x^{2}}=\frac{1}{k_{p}t}\frac{1+x}{x}\left((1+x)^{2}+2\frac{k_{p}}{k_{\text{off}}}\right).\label{delta-x-from-n}
\end{equation}
For $N$ independent copies of the receptor on the cell surface, the above expression is multiplied by $N^{-1}$ \citep{Bialek2005}.

These results are summarized in Figs. \ref{fig:mode_1_fig}(d) and \ref{fig:mode_1_fig}(e), which show the squared relative error of the estimate $x$ (equivalently, $k_{\text{off}}$),
scaled by $k_{p}t/N$, as a function of dimensionless quantities $k_{\text{on}}c/k_{p}$ and $k_{\text{off}}/k_{p}$. 
We scale the squared relative error by $k_{p} t/N$ to present the time-independent part of the cell's estimation error. 
The blue region in Fig. \ref{fig:mode_1_fig}(d) indicates where the error in the estimate (i.e. $\sqrt{\langle\delta x^2 \rangle/N}$) is less than 33\% the true value, which we have chosen as a cutoff for ``good'' estimation performance. This cutoff is chosen somewhat arbitrarily, as the cell can improve its estimation accuracy by increasing the number of copies $N$ of the receptor or by increasing the signal integration time $t$.
To provide a conservative estimate, we visualize these results with $N=10^2$, which is at the lower end of the biological range
of receptor expression levels \cite{Piehler2012}.

As expected, the typical error of the estimate of $k_{\text{off}}$ diverges both for
$x\rightarrow0$ and $x\rightarrow\infty$ because at very fast unbinding rates the receptor does not produce enough signaling molecules
for a meaningful statistical estimate, while at very slow unbinding rates
the receptor occupancy saturates independent of either affinity or
concentration. The value of $x$ at which the optimal specificity is achieved increases with the concentration $c$ and saturates to $x=1/2$ for $c\gg 1$.

The problem of concentration sensing accuracy can be similarly formulated as the problem
of estimating $c$ from $n$ at fixed $k_{\text{off}}$ via maximization
of the likelihood $P(n|c)$ over $c$ \citep{Endres2009}. Following
the same approach as in Eqs. (\ref{eq-c-estimate-simple-receptor})
and (\ref{delta-x-from-n}), the best estimate and its variance are
again given by $\frac{k_{\text{on}}c^{*}}{k_{\text{off}}}=\frac{n}{k_{p}t}\frac{1}{1-n/k_{p}t}$
and $\frac{\langle\delta c^{2}\rangle}{c^{2}}=\frac{\langle \delta x^{2}\rangle}{x^{2}}=\frac{\langle\delta k_{\text{off}}^{2}\rangle}{k_{\text{off}}^{2}}$.

Interestingly, the ML estimate of the concentration based on $n$ is mathematically identical to that of Endres and Wingreen
\citep{Endres2009}, which was based on a more informative likelihood function
containing the whole series of binding and unbinding events. We return
to this point in the Discussion.

In the limit $k_{p}\gg k_{\text{off}}$ the expression for the concentration sensing accuracy
reduces to the classical Berg-Purcell expression $2(1+x)/(k_{\text{on}}ct)=1/(2Dac(1-p)t)$
where $4Da=k_{\text{on}}$ \citep{Berg1977,TenWolde2016}, as each binding
event in this regime produces multiple signaling molecules, and the
sensing accuracy is limited by the fluctuations in receptor occupancy considered in \citep{Berg1977}.
Notably, for finite $k_{p}/k_{\text{off}}$, fluctuations in the production
of the output molecule $n$ play an important role, as they cause
the divergence of the estimate accuracy observed at high $x$ in Eq.~\ref{delta-x-from-n} - a feature absent
from the models that consider only the noise in the receptor occupancy \citep{Berg1977,Endres2009}.
In particular, the concentration at which the best estimate is obtained
changes significantly with $k_{p}/k_{\text{off}}$, and is not necessarily
close to $x=1$, the point of highest response sensitivity to concentration
changes.

However, crucially for the main question of this paper, it is impossible to estimate both $c$ and $k_{\text{off}}$ simultaneously because the distribution $P(n|c,k_{\text{off}})$ does
not possess a well defined peak in the $(c,k_{\text{off}})$ space
but rather a ridge along the line $\frac{c}{K_{d}}=\frac{n}{k_{p}t}\frac{1}{1-n/k_{p}t}$,
as shown in Fig. \ref{fig:mode_1_fig}(f). This could in principle
be resolved by careful selection of a prior on $(c,k_{\text{off}})$,
which in practice implies additional assumptions regarding the molecular
inference machinery, and lies outside the scope of the present work.

\subsection{Pleiotropic receptor}
\label{subsec:pleiotropic-receptor}

\begin{figure*}[!h]
\centering
\includegraphics[width=13 cm]{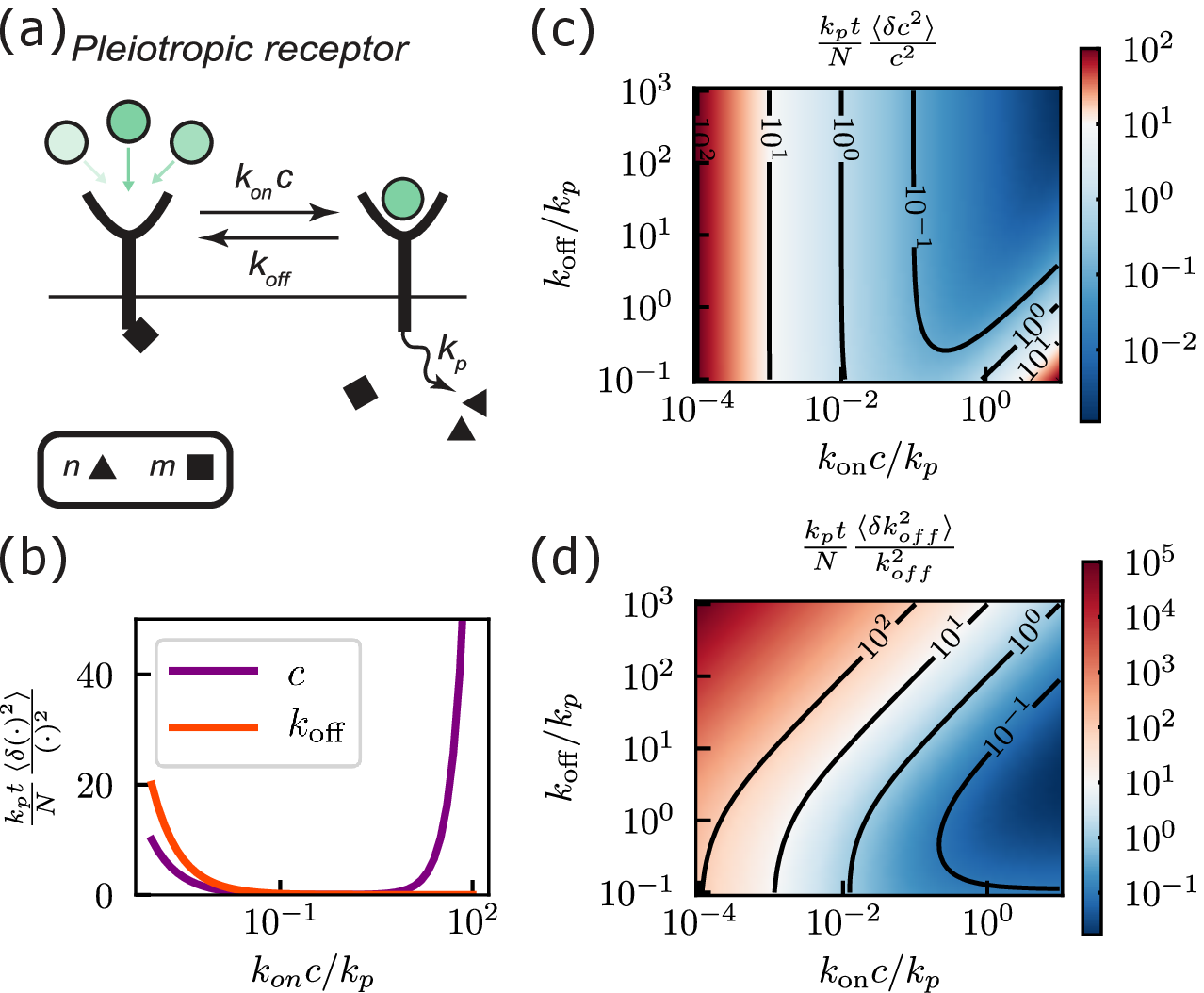}
\caption{\textit{Specificity and accuracy of pleiotropic signaling.} (a) The pleiotropic receptor considered
here has the same behaviour as in Fig. \ref{fig:mode_1_fig}, except
that an additional output molecule, denoted as $m$, is produced for
each binding event. (b) The scaled squared relative errors of the
estimates for $k_{\text{off}}$ (orange) and $c$ (purple) as a function
of $k_{\text{on}}c/k_{p}$, holding $k_{\text{off}}/k_{p}=1$ and $ k_{p} t = 10^2 $; see
Eq. (\ref{model2-err}). (c) and (d) Scaled squared relative estimate errors $ \frac{k_{p} t}{N} \frac{ \langle \delta k_{\text{off}}^2 \rangle }{ k_{\text{off}}^2 }$ and $\frac{k_{p} t}{N} \frac{ \langle \delta c^2 \rangle}{c^2} $, for $ k_{p} t = 10^2$ and $ N = 10^2 $.}
\label{fig:mode_2_fig}
%\end
\end{figure*}

In contrast to the results of the previous section, it is possible
to unambiguously estimate both $k_{\text{off}}$ and $c$ simultaneously
if the receptor is pleiotropic (i.e. produces more than one output
signal upon ligand binding). In this section, we extend the model
of the previous section to include a second sensing molecule, inspired
by G-protein coupled receptor (GPCR) signaling \citep{neves2002g,Wootten2018},
shown in Fig. \ref{fig:mode_2_fig}(a). 
In this scheme, the G-protein
(GP)-like molecule is pre-bound to the intracellular domain of the
receptor and detaches once a ligand binds. For simplicity, we assume
that upon ligand unbinding the receptor quickly rebinds a new GP-like
molecule and is ready for signaling. Although, as defined, this type of sensing molecule has
an unnatural feature that it can be produced even for infinitesimally
short binding events \citep{Carballo-Pacheco2019}, the model is sufficient
to demonstrate the role of pleiotropy in ligand sensing.
Biologically, combined kinase-phosphorylation
and G-protein signaling has been reported in some cytokine receptors
\citep{Rawlings2004} and other immune receptors \citep{mocsai2010syk}. 

We denote the total number of GP-like molecules produced by time $t$
as $m$, which effectively serves as a count of the number of binding
events rather than the total binding time (which is measured by $n$, as defined
in the previous section) \citep{Endres2009,Mora2015,Singh2017}. As shown below, this
model naturally allows joint determination of the ligand identity and its quantity
$(c,k_{\text{off}})$ from the two signaling outputs $(n,m)$ (see
Fig. \ref{fig:mode_2_fig}).

%\begin{SCfigure*}[\sidecaptionrelwidth][t]

The system is now described by the probability of being in a given
state at time $t$, $P_{i}^{n,m}(t)$, where $i$ denotes the receptor
state ($i=1$ if bound by a ligand, $0$ if unbound) and $n$ and
$m$ are the numbers of the sensing molecules accumulated by time $t$. 
Similar to the previous section, the dynamics of the system are described by the following master
equation:
\begin{equation}
\begin{aligned}\frac{d}{dt}P_{0}^{n,m} & =k_{\text{off}}P_{1}^{n,m}-k_{\text{on}}cP_{0}^{n,m}\\
\frac{d}{dt}P_{1}^{n,m} & =k_{\text{on}}cP_{0}^{n,m-1}-k_{\text{off}}P_{1}^{n,m}+k_{p}P_{1}^{n-1,m}-k_{p}P_{1}^{n,m}.
\end{aligned}
\label{eq-ME-pleiotropy}
\end{equation}

This master equation can be solved using the generating function technique,
similar to Eq. (\ref{gen-fun-eq}) in the previous section (see
details in the Supplemental Material \citep{SM2020}). The mean and the variance of $n$ remain the same
as in Eq. (\ref{nonpleio_meann}). The mean of $m$, its variance
$\langle{\delta m}^{2}\rangle$ and the covariance $\langle{\delta n}{\delta m}\rangle$
are, in the $k_{\text{off}}t\gg1$ limit,
\begin{equation} \label{pleio_meanm_meann}
\begin{aligned}\langle m\rangle & =k_{\text{off}}t\frac{x}{1+x}\\
\langle{\delta m}^{2}\rangle & =k_{\text{off}}tx\frac{1+x^{2}}{(1+x)^{3}}\\
\langle{\delta n}{\delta m}\rangle & =k_{p}tx\frac{1-x}{(1+x)^{3}}.
\end{aligned}
\end{equation}

These results can also be derived using renewal process theory
(see Supplemental Material \citep{SM2020}). Note that at small $x$, $n$ and $m$ are correlated because
in the low concentration/weak binding limit the overall bound time
is proportional to the number of events. By contrast, at large $x$,
$n$ and $m$ are anti-correlated because in this regime a time series
with more binding-unbinding events results in lower overall bound
time. However, for $x\rightarrow\infty$, $\langle{\delta n}{\delta m}\rangle\rightarrow0$
because the receptor is occupied all the time, and the number of binding
events is not correlated with the total bound time.

\textbf {\emph{Specificity and accuracy of the pleiotropic receptor}.} The crucial feature of Eq. (\ref{pleio_meanm_meann}) is that the variable $m$ depends differently on the unbinding rate $k_{\text{off}}$ and the concentration $c$ compared to the variable
$n$ of Eq. (\ref{nonpleio_meann}), which allows
the estimation of both $c$ and $k_{\text{off}}$. As before, we assume
that $k_{p}$ and $k_{\text{on}}$ are fixed constants, hardwired
into the molecular machinery of the cell. In the long time limit,
the likelihood $P(n,m|c,k_{\text{off}})\equiv P_{0}^{n,m}+P_{1}^{n,m}$
is well approximated by a multivariate Normal distribution $\mathcal{N}(\boldsymbol{\upmu},\pmb{\hat{\text{C}}})$
with mean and covariance
\begin{equation}\label{multivariate_gaussian}
\boldsymbol{\upmu}=\begin{bmatrix}\langle n\rangle\\
\langle m\rangle
\end{bmatrix}\quad\text{and}\quad\pmb{\hat{\text{C}}}=\begin{bmatrix}\langle{\delta n}^{2}\rangle & \langle{\delta n}{\delta m}\rangle\\
\langle{\delta m}{\delta n}\rangle & \langle{\delta m}^{2}\rangle
\end{bmatrix},
\end{equation}
so that 
\begin{equation*}
P(n,m|c,k_{\text{off}})=(Z)^{-1}  \exp\left(-\frac{1}{2}(\textbf{n}-\boldsymbol{\upmu})^{T}\pmb{\hat{\text{C}}}^{-1}(\textbf{n}-\boldsymbol{\upmu})\right), 
\end{equation*}
where $\textbf{n}=(n,m)$ and normalization factor $Z=(2\pi)^2\det(\pmb{\hat{\text{C}}})^{1/2}$. 

Estimates for $k_{\text{off}}$ and $c$ can be found in the same
manner as in the previous section, by maximizing the likelihood $P(n,m|c,k_{\text{off}})$
over $c$ and $k_{\text{off}}$, which yields:
\begin{equation}
c^{*}=\frac{k_{\text{off}}^{*}}{k_{\text{on}}}\frac{n/(k_{p}t)}{1-n/k_{p}t}\;\;\;\;k_{\text{off}}^{*}=k_{p}\frac{m}{n}.
\label{eq:MLE-est-pleio}
\end{equation}

Interestingly, the same ML estimates are obtained using a more
detailed likelihood of binding and unbinding times introduced in \citep{Endres2009}, indicating that the signaling scheme studied here might be close to optimal. Despite this similarity, we emphasize that \citep{Endres2009} did not address the question of specificity, being focused on the optimality of the concentration sensing.

In a generalization of the one-variable procedure from section \ref{subsec:non-pleio}, the expected estimate uncertainties are given by the typical widths of the posterior/likelihood in the $(c,k_{\text{off}})$ space, quantified by the covariance matrix
\begin{align}\nonumber
\pmb{\hat{\Sigma}}&\equiv\left[\begin{array}{cc}
\langle\delta c^{2}\rangle & \langle\delta k_{\text{off}}\delta c\rangle\\
\langle\delta k_{\text{off}}\delta c\rangle & \langle\delta k_{\text{off}}^{2}\rangle
\end{array}\right].\nonumber
\end{align}

Very generally, a lower bound on the covariance matrix of the estimates is given by the inverse of the Fisher Information Matrix (FIM) $\pmb{\hat{\mathcal{I}}}$
\begin{equation*}
\pmb{\hat{\mathcal{I}}}(c, k_\text{off})\equiv
-\left[\begin{array}{cc}
\langle\frac{\partial^2 L}{\partial c \partial c}\rangle & \langle\frac{\partial^2 L}{\partial k_\text{off} \partial c}\rangle \\
\langle\frac{\partial^2 L}{\partial c \partial k_\text{off}} \rangle& \langle\frac{\partial^2 L}{\partial k_\text{off} \partial k_\text{off}}\rangle
\end{array}\right],
\end{equation*}
where $L$ is the logarithm of the likelihood $P(n,m|c,k_{\text{off}})$ \citep{KayTextCh3}.

In the long time limit, where the likelihood is sharply peaked, the approximate squared relative errors of the estimates are
\begin{equation}
\begin{aligned}\frac{\langle\delta c^{2}\rangle}{c^{2}} & =\frac{1}{k_{p}t}\frac{1+x}{x}\left(x^{2}+\frac{k_{p}}{k_{\text{off}}}\right)\\
\frac{\langle\delta k_{\text{off}}^{2}\rangle}{k_{\text{off}}^{2}} & =\frac{1}{k_{p}t}\frac{1+x}{x}\left(1+\frac{k_{p}}{k_{\text{off}}}\right).\label{model2-err}
\end{aligned}
\end{equation}

The scaled squared relative errors are plotted in Fig. \ref{fig:mode_2_fig}.
In plain language, the cell is not capable of distinguishing
ligands with affinities closer than $\sqrt{\langle\delta k_{\text{off}}^{2}\rangle/N}$. An
important consequence of this analysis is that signaling specificity
is not determined solely by the differences between ligand affinities but also depends on the
ligand concentrations, and is thus context-dependent. The relative
error of the concentration estimate behaves qualitatively similar
to that of the non-pleiotropic receptor, diverging in both the low
and high $x$ limits. On the other hand, the error in $k_{\text{off}}$
inference remains low even in the high-occupancy (high $x$) regime
because the knowledge of both $n$ and $m$ allows accurate determination
of the average bound time \citep{Endres2009,Singh2020}.

The ratio of the concentration sensing error
of the pleiotropic receptor (Eq. (\ref{model2-err}))
to that of the non-pleiotropic receptor (Eq. (\ref{delta-x-from-n}))
is $(x^{2}+k_{p}/k_{\text{off}})/((1+x)^{2}+2k_{p}/k_{\text{off}})$.
Since this quantity ranges between $(k_{p}/k_{\text{off}})/(1+2k_{p}/k_{\text{off}})$
at low $x$ and $1$ at high $x$, pleiotropy always improves the
concentration sensing accuracy. Likewise, the corresponding ratio of the estimator errors
for $k_{\text{off}}$ is always less than one,
indicating that pleiotropy not only enables simultaneous estimation of the concentration and the affinity but generally also increases the
specificity of signaling as well. This improvement is non-trivial because our pleiotropic model must estimate two variables, rather than just one variable in the non-pleiotropic case. 

\subsection{Discrimination between multiple ligands in a mixture}
\label{subsec:ligands-mixture}
	
We now consider the more challenging problem of cross talk in a simultaneous mixture of multiple  ligands distinguished by their affinities $k_{\text{off,1}}$ and $k_{\text{off,2}}$.
This problem is common in many signaling contexts such as cytokine and TCR signaling in the immune system, as well as others (e.g. \citep{Moraga2014,Singh2013,Lee2007,Antebi2017,Feng2005,Shuai2003}).
In this section, extending the minimal scheme introduced above, we demonstrate that even in this more challenging case both the identities and concentrations of each ligand can be estimated well based on the pleiotropic receptor outputs.

\begin{figure*}[!h]
\centering
\includegraphics[width=16 cm]{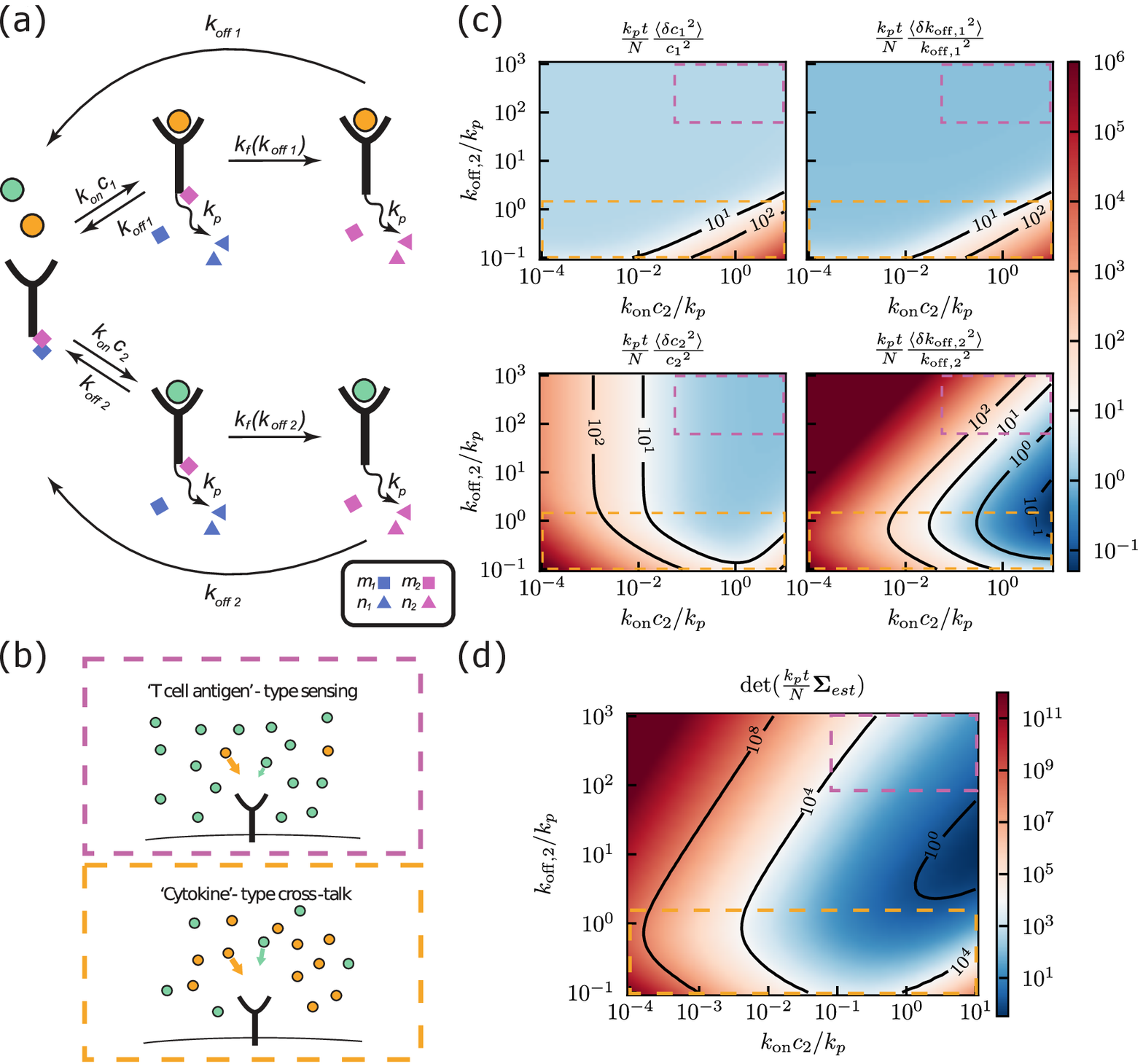}
\caption{\textit{Schematic of sensing multiple ligands}. 
(a) The cross talk receptor considered here produces four distinct sensing molecules when bound. Upon ligand binding, the receptor enters the first bound state from which it releases one $m_1$ molecule and produces $n_1$ molecules at rate $k_p$, analogously to the pleiotropic receptor. Depending on the identity of the bound ligand (encoded through $k_{\text{off}}$), the receptor can either unbind or transition to a second bound state with rate $k_f = \alpha/k_{\text{off}}$, from which it releases one $m_2$ molecule and produces $n_2$ molecules with rate $k_p$. 
(b) One cross talk signaling challenge (e.g. cytokines) is to sense two ligands with similar affinities present at arbitrary concentrations. A second cross talk challenge (e.g. T-cell antigens) is to accurately sense a high affinity ligand (large arrow) present at low concentrations while simultaneously sensing a low affinity ligand (small arrow) present at high concentrations. 
(c) Heatmaps of the diagonal elements of the normalized estimate covariance matrix, scaled by $k_p t/N$. The affinity and concentration of the stronger binding ligand are fixed and the affinity and concentration of the weaker binding ligand are varied. 
(d) Corresponding heatmap of the determinant of the normalized estimate covariance matrix. The orange (purple) dashed rectangle indicates the region corresponding to the `cytokine' (`T-cell antigen') type challenge illustrated in (b). Parameters used are $\alpha/{k_{p}}^2=0.2$, $k_{p}t = 10^{2}$, $N = 10^{2}$, $k_{\text{on}}c_1/k_{p}=10^{-1}$ and $k_{\text{off,1}}/k_{p}=10^{-2}$.
}
\label{fig:multilig-fig}
%\end{SCfigure*}
\end{figure*}

For simplicity, we confine ourselves here to a mixture of two ligands. Following the approach of the previous section, we assume that the goal of the cell is to infer the affinity $k_{\text{off, i}}$ and the concentration $c_i$ of each ligand in the mixture. As there are four quantities to infer, we consider an extension of the scheme of Fig. \ref{fig:mode_2_fig}(a) whereby four different signaling molecules are produced by the receptor in response to ligand binding. To achieve this we assume that the receptor can be in two different bound states (for each ligand) as seen in Fig. \ref{fig:multilig-fig}(a). Molecularly, these states can be different receptor conformations, different phosphorylation states, or have different binding co-factors present \citep{Lodish2000}.

As in the previous section, we assume that the binding of either ligand immediately induces the release of an $m$-type molecule $m_1$ from the receptor; and an $n$-type sensing molecule $n_1$ is produced with rate $k_p$ while the receptor remains in this first bound state.
Transition to the second state occurs with the rate $k_f$, and releases a different sensing molecules $m_2$; accordingly, $n_2$ molecules are produced (also with rate $k_p$) while the receptor is in the second state.
From either of these bound states, the ligand can unbind with the rate $k_{\text{off,i}}$.
For simplicity, we follow the common assumption that the transitions from first state to the second are irreversible.
Although commonly assumed to be constant, in general $k_f$ may depend on the ligand identity \citep{savir2007conformational}, because the energies of both states depend on the ligand binding affinity. We choose the transition rate to be inversely proportional to the affinity $k_f=\alpha/k_{\text{off}}$ where $\alpha$ is a constant (see Supplemental Material \citep{SM2020}); this assumption does not affect the main results of the paper, and can be relaxed.
Note that although the resulting kinetic scheme resembles the well-known kinetic proofreading (KPR) scheme \citep{McKeithan1995,fehervari2019proving}, this is incidental to our results, and it is employed here as a simple model of a multi-state receptor. KPR-type approaches to the cross talk problem will be investigated in more detail elsewhere.

%================================================Fig. 3
%\begin{SCfigure*}[\sidecaptionrelwidth][t]

%================================================Fig. 3

The dynamics of the distribution $P(n_1,m_1,n_2,m_2|c_1,k_{\text{off},1},c_2,k_{\text{off},2})$ are described by an appropriate master equation, similarly to the previous section (see Supplemental Material \citep{SM2020} for details). For sufficiently long times it is also well approximated by a Normal distribution similar to Eq. (\ref{multivariate_gaussian}).
The means of $n_1$, $m_1$, $n_2$, and $m_2$ are given by
\begin{equation}
\begin{aligned}
    \langle n_1 \rangle &= \frac{k_{p} t}{1+x_1+x_2} \left( \frac{x_1}{1+\alpha/k_{\text{off,1}}^2} + \frac{x_2}{1+\alpha/k_{\text{off,2}}^2} \right)\\
    \langle m_1 \rangle &= k_{\text{on}} t \frac{c_1 + c_2}{1+x_1+x_2} \\
    \langle n_2 \rangle &= \frac{k_{p} t}{1+x_1+x_2} \left( \frac{x_1}{1+k_{\text{off,1}}^2/\alpha} + \frac{x_2}{1+k_{\text{off,2}}^2/\alpha} \right) \\
    \langle m_2 \rangle &= \frac{k_{\text{on}} t}{1+x_1+x_2} \left( \frac{c_1}{1 + k_{\text{off,1}}^2/\alpha} + \frac{c_2}{1+k_{\text{off,2}}^2/\alpha}\right)\,. 
\end{aligned}
\end{equation}
The explicit expressions for the second moments and the covariances are more complicated, and are presented in the Supplemental Material \citep{SM2020}.

Following the same prescription as in the previous section, the estimates of the ligand identities and their concentrations, $c_1,k_{\text{off},1},c_2,k_{\text{off},2}$ can be obtained based on the measurements of the output variables $n_1,m_1,n_2,m_2$ using, for instance, a Maximum Likelihood estimator. 
As before, the lower bounds on the estimate errors are provided by the elements of the Fisher Information Matrix. 
In this section, we restrict ourselves to a numeric analysis of the FIM due to the added complexity of the resulting expressions compared to the expressions for the single ligand case (see Supplemental Material \citep{SM2020} for details). 

The results are shown in Fig. \ref{fig:multilig-fig}. The diagonal elements of $\pmb{\hat{\mathcal{I}}}^{-1}$ shown in Fig. \ref{fig:multilig-fig}(c) provide approximate lower bounds on the mean errors of the estimates of the affinities and the concentrations of both ligands.
In addition, Fig. \ref{fig:multilig-fig}(d) shows the determinant of the inverse FIM, $\text{det}(\pmb{\hat{\mathcal{I}}}^{-1})$, which serves as a global measure of the overall estimation error \citep{Nemenman2010,Carballo-Pacheco2019}.
The blue region in Fig. \ref{fig:multilig-fig}(c) indicates where the error in the estimate is less than 33\% of the true value. Accordingly, the blue region in Fig. \ref{fig:multilig-fig}(d) indicates where the overall estimation has low error. 
As $k_{\text{off,2}}$ approaches $k_{\text{off,1}}$ the estimation problem becomes ill-posed (i.e. $\pmb{\hat{\mathcal{I}}}$ becomes singular). In this limit, the receptor outputs are no longer independent and so four input variables cannot be estimated simultaneously.

It is instructive to focus on two different regions of the parameter space in Figs.
\ref{fig:multilig-fig}(c) and \ref{fig:multilig-fig}(d)
which correspond to two well-known sensing problems commonly encountered in a number of cell signaling contexts.
One problem, illustrated in the lower part of Fig. \ref{fig:multilig-fig}(b) (orange dashed line), is the specific and accurate sensing of two ligands in a mixture when both have similar affinities and arbitrary concentrations. This type of challenge is commonly present, for example, in cytokine and chemokine signaling in the immune system \citep{Moraga2014,Sokol2015}.
This scenario corresponds to the lower regions of the plots in Figs. \ref{fig:multilig-fig}(c) and \ref{fig:multilig-fig}(d), indicated by the orange dashed rectangle (the weaker binding ligand has an affinity that is relatively close to that of the stronger binding ligand).
As shown in Fig. \ref{fig:multilig-fig}(c), this is a difficult task to perform, and is possible only in an intermediate range of concentrations where the blue regions in each panel overlap (roughly between $k_{on}c_2/k_p \in [10^{-2},10^0]$, or within the $10^4$ contour of Fig. \ref{fig:multilig-fig}(d)).
This finding might be part of an explanation of the prevalence of dimeric receptors in many signaling pathways, because ligand induced dimerization enhances the differences in the effective binding affinities between similar ligands \citep{Fathi2016}.
Overall, there is a trade-off between the two ligands: a specific estimate of $k_{\text{off,1}}$ implies a less specific estimate of $k_{\text{off,2}}$; similar trade-off exists for accuracy of the estimation of concentrations $c_1$ and $c_2$.

The ``T cell antigen'' scenario is illustrated in the upper region of Fig. \ref{fig:multilig-fig}(b) (purple dashed line). 
It represents a different sensing problem, whereby the stronger binding ligand is present at a low concentration on the background of a low affinity ligand present at a high concentration \citep{Francois2013,Francois2016,Mora2015}. 
This scenario corresponds to the upper right region in Figs. \ref{fig:multilig-fig}(c) and \ref{fig:multilig-fig}(d) (indicated by a purple dashed line). The plots of the accuracy and specificity for ligand 1 (two upper panels in Fig. \ref{fig:multilig-fig}(c)) indicate that for sufficiently distinct ligands (i.e. $k_{\text{off}, 2} \gg k_{\text{off}, 1}$), sensing of the stronger-binding ligand is not impaired by a wide range of concentrations of the weaker-binding ligand. Interestingly, in this regime overall estimation has low error (blue region of Fig. \ref{fig:multilig-fig}(d)).  We discuss the implications of these results in the next section.

\section{Summary and Discussion}
\label{sec:sum-dis}

Cross talk is common in many signaling pathways, which raises the question of how cells are able to sense and thereby respond appropriately to molecularly similar signals carrying different information through these cross-wired pathways. In this paper, we focused on cross talk at the ligand-receptor level, whereby multiple ligands can act through
the same surface receptor. This situation can be encountered
in cytokine and chemokine signaling in the immune system \citep{Rowland2012,Moraga2014,Sokol2015},
developmental pathways \citep{Touzot2014,Feng2005,Tran2018,Antebi2017}
and other physiological systems \citep{zinkle2018threshold,qin2019optimal}. 
Ligand-receptor cross talk entails a fundamental problem: it is impossible
to discriminate between different cognate ligands based on receptor
occupancy alone because the identity of the ligand (``quality'')
can be confounded by its concentration (``quantity'') (e.g. \citep{Francois2016,francois2019physical}). 
Equally important, when multiple different ligands can bind to the same receptor, it is unclear how the downstream signaling machinery can distinguish between various combinations of different ligands based on the receptor activity alone. 

In this paper, we have investigated one potential solution to this
problem - signaling pleiotropy - which commonly accompanies cross talk
\citep{Ozaki2002,Moraga2014}, using models of
receptor kinetics that account for the molecular noise at both the
receptor and the downstream variables. 
We mathematically confirmed the intuitive notion that the classical model of a receptor which binds multiple ligands but produces only one type of downstream sensing molecule is not able to simultaneously discern the ligand identity (as defined by its unbinding rate $k_{\text{off}}$), and its quantity as expressed by its concentration $c$. 

In contrast, a pleiotropic receptor, which produces two types of downstream signaling molecules, can resolve this ambiguity. 
The crucial feature of the model enabling these properties is that the two output
signaling molecules reflect physically different features of the ligand-receptor interaction - in the case studied here, one variable is proportional to the bound time of the ligand, while the other reflects the number of distinct ligand-receptor binding events. 

Importantly, a realistic feature of our model is that the inference is based only on the numbers of the produced signaling molecules and not on the knowledge of whole sequence of binding-unbinding events that has been a feature of a number of works \citep{Endres2009,Singh2017,Mora2010,Singh2020}. In addition to providing a solution to the specificity-accuracy dilemma, our analysis indicates that the noise in the production of the downstream sensing molecules (on top of the receptor binding-unbinding fluctuations) can significantly affect the specificity and the accuracy of molecular sensing compared to the models that
only account for the randomness of the receptor-unbinding events \citep{Endres2009,Singh2020,TenWolde2016}.

Furthermore, we have shown that with a sufficient number of downstream receptor sensing molecules, a pleiotropic receptor can simultaneously determine the identities and quantify the concentrations of two distinct (and arbitrary) ligands, even when they are present in arbitrary mixtures.
Namely, four output variables can be used to determine the four unknown variables - two dissociation constants and two concentrations of the ligands in the mixture.
Due to the robust nature of this discrimination scheme, we expect it to generalize to combinations of multiple ligands, whereby a signaling network with $2L$ output molecules can be used to determine the identities and the concentrations of $L$ ligands in the mixture.
One such potential scheme is a KPR-like signaling chain with $2L$ bound states of the receptor.
This could provide a foundation for a generic combinatorial signal recognition system in cross-wired pathways.
The estimation errors are likely to increase with the number of ligands in the mixture \citep{Singh2020}, and it remains to be investigated up to what number of ligands the proposed mechanism remains biologically relevant.
However, the error can always be lowered by increasing the copy number of receptors or the signaling time.
Investigation of how the sensing error scales with the number of ligands is beyond the scope of this manuscript and will be presented elsewhere.

Due to the interplay of molecular and evolutionary constraints, cross talk in receptor signaling systems may be inevitable, and our results indicate that receptor pleiotropy can provide a general mechanism for specific and accurate signaling in such systems.
However, signaling cross talk might not be just undesirable ``noise'' hampering accurate and specific ligand discrimination, but rather might have benefits of its own.
For example, the combination of cross talk and pleiotropy may require fewer types of receptors than multiple independent receptors for each ligand, which could lead to lower resource requirements.
Additionally, pleiotropic cross-wired signaling networks may be more robust to genetic or structural alterations in the molecular components of the pathway such as the receptors themselves, or the associated the readout molecules and the signaling enzymes.

Our model is constrained by the assumptions about the nature of the
output variables $n$ and $m$, which are inspired by the observed
modes of signaling in cytokine and GPCR pathways.
It is instructive to compare our results with those based on the inference from the whole sequence of binding-unbinding events (which requires more intricate intracellular molecular networks) \citep{Endres2009,Mora2015,Singh2020,Singh2017,Siggia2013}.
The likelihood of a sequence of binding/unbinding events with overall bound time $t_{b}$ and the overall number of binding events $m$ is given by $P(t_{b},m|c,k_{\text{off}})\sim\exp(-k_\text{on}ct)\exp(t_{b}(-k_{\text{off}}+k_{\text{on}}c))\cdot(k_{\text{off}}k_{\text{on}}c)^{m}$
\citep{Endres2009}.
Maximizing this likelihood over $(c,k_{\text{off}})$
results in the same estimates as given by Eq. (\ref{eq:MLE-est-pleio}).
The corresponding lower bounds on the estimation errors (found by
inverting the Fisher Information Matrix) in this case are $\langle\delta c^{2}\rangle/c^{2}=\langle\delta k_{\text{off}}^{2}\rangle/k_{\text{off}}^2=1/\langle m\rangle$,
where $\langle m\rangle=k_{\text{off}}tx/(1+x)$. These expressions
match Eq. (\ref{model2-err}) in the limit $k_{p}/k_{\text{off}}\gg1$ (at
finite $x)$; the deviation at finite $k_{p}/k_{\text{off}}$ is a
consequence of the additional noise in the production of the signaling
molecules on top of the noise of receptor-ligand binding. Thus, our minimal model with only two readout variables appears to be able to take advantage of the full information encoded in the whole sequence of binding-unbinding events. This is likely due to the fact that the Endres-Wingreen likelihood effectively only depends on $t_b$ and $m$ - which are the physical variables ``measured'' by our variables $n$ and $m$ \citep{Endres2009,Siggia2013}. This is not true anymore for multiple ligands in a mixture where the likelihood of the time series of the binding-unbinding events retains the explicit dependence on the specific sequence of bound times \citep{Mora2015,Siggia2013}. In the future, it will be important to investigate what molecular mechanisms might enable cells to access the additional information contained in the sequence of binding-unbinding events \cite{Singh2020}.

In this paper we have considered only a minimal ``module''
of a cross-wired receptor signaling network - a receptor of one type
capable of interacting with multiple types of ligands. Furthermore,
we have assumed that the cell performs estimates of ligand identity and concentration based on the immediate downstream receptor outputs. The
ideas of this paper can be extended to the more general case of complex
networks of ligands, receptors and downstream signaling molecules
that may include positive and negative feedbacks mediated by such molecules
\citep{Francois2013,Murugan2012,Singh2020,Carballo-Pacheco2019}. The pleiotropic receptor outputs we considered here may also be used for alternative sensing goals such as detecting changes in the composition of a ligand mixture \citep{Siggia2013}.
Functional pleiotropy can alternatively be achieved by using different stages in the time course of the signaling as distinct output variables
\citep{Jetka2018,freed2017egfr,Siggia2013}, facilitated through feedback such as receptor internalization.
%Indeed, molecular encodings of arbitrary moments of the full distribution of binding and unbinding events may provide a %general way to construct pleiotropic receptor outputs \citep{Singh2020}.

Depending on the task that needs to be solved by the cell, different mathematical frameworks might be more appropriate \cite{Siggia2013}. Among other approaches, the results of this paper also provide
an interesting outlook on the information theory approaches to cell
signaling \citep{Jetka2018,Ellison2016,Tkacik2009, waltermann2011information, mehta2009information,barato2014efficiency} via connections between the channel capacity of a signaling pathway and the Fisher Information Matrix. Further constraints, such as receptor integration time and energy consumption in the non-equilibrium signaling cycle \citep{Govern2014,MehtaMora2014} may be considered in extensions of this work. 
These lie outside of the scope of the present work, and will be studied in the future.
Finally, while 
this paper has focused on sensing capabilities of single cells, inter-cellular
interactions in multicellular environments can result in collective
responses to the signaling milieu leading to collective decisions
at the population level \citep{Mugler2016,oyler2017tunable,Suderman2017}.

\section*{Materials and Methods}
All calculations and simulations were performed in Python 3.8.3 using NumPy 1.18.5, and Mathematica 10.4. Computational details of the results presented here are provided in the Supplemental Material. The code implementing the analysis, simulations, and plotting is available at \url{https://github.com/uoftbiophysics/cellsignalling}. \\

\section*{Supplemental materials}
\noindent Supplemental Text\\
section A: Calculating the moments of the output variables\\
section B: Obtaining Maximum Likelihood Estimates of $c$ and $k_\text{off}$\\
section C: Accuracy of the estimators\\
section D: Alternative calculation of the moments of the distribution of $n$ and $m$\\
section E: Long time distribution well-approximated by a Normal distribution\\
section F: Bayesian framework, effects of prior\\
fig. S1. The MLE for $x$ as a function of $n$\\
fig. S2. Model 2 MLE as a function of $n$ for fixed $m$\\
fig. S3. Comparison of heuristic estimation error and the Cram\'{e}r-Rao lower bound for Model 2\\
fig. S4. Long-time distributions of $n$ and $m$\\
fig. S5. Estimation with prior\\
references \cite{Kampen2007,Freeman1997,Ash2000}

\textbf{Acknowledgements.} The authors are indebted to numerous colleagues in the field for illuminating discussions. 
\textbf{Funding:} The authors acknowledge the support of the
Natural Sciences and Engineering Research Council of Canada (NSERC) through Discovery Grant RGPIN 402591 to A.Z.; CGS-D Graduate Fellowship to M.S. and PGS-D Graduate Fellowship to D.K.\\
\textbf{Contributions:} D.K., J.R., and M.S. contributed equally to this work. A.Z. devised the project. All authors performed the research and wrote the paper.\\
\textbf{Competing interests:} The authors declare that they have no competing interests.

% Bibliography
%\citep{Kampen2007,Freeman1997}
%\bibliography{receptor-signaling-july6-az}
%\nocite{Kampen2007}
%\nocite{Freeman1997}
%\nocite{Ash2000}

%

\end{document}